\documentclass[11pt]{article}
\usepackage{amssymb}
\usepackage{latexsym}
\begin{document}
\setlength{\baselineskip}{15pt}
\title{Congruence method for global Darboux reduction \\
of finite-dimensional Poisson systems}
\author{ \mbox{} \\ Benito Hern\'{a}ndez-Bermejo $^1$}
\date{}

\maketitle
\begin{center}
{\em Departamento de Biolog\'{\i}a y Geolog\'{\i}a, F\'{\i}sica y Qu\'{\i}mica Inorg\'{a}nica. \\ 
Universidad Rey Juan Carlos. Calle Tulip\'{a}n S/N. 28933--M\'{o}stoles--Madrid. Spain.} 
\end{center}

\mbox{}

\mbox{}

\mbox{}

\begin{center} 
{\bf Abstract}
\end{center}
\noindent
A new procedure for the global construction of the Casimir invariants and Darboux canonical form for finite-dimensional Poisson systems is developed. This approach is based on the concept of matrix congruence, and can be applied without the previous determination of the Casimir invariants (recall that their prior knowledge is unavoidable for the standard reduction methods, thus requiring either the integration of a system of PDEs or solving some equivalent problem). Well the opposite, in the new congruence method both the Darboux coordinates and the Casimir invariants arise simultaeously as the outcome of the reduction algorithm. In fact, the congruence algorithm proceeds only in terms of matrix-algebraic transformations and direct quadratures, thus avoiding the need of previously integrating a system of PDEs and therefore improving previously known approaches. Physical examples illustrating different aspects of the theory are provided. 

% ABSTRACT PARA PHYSICS LETTERS A
%A new procedure for the global construction of the Casimir invariants and the Darboux canonical form is provided. Such method is %algorithmic, algebraic, of simple application and often improves previously known approaches. Examples are given. 

\mbox{}

\mbox{}
 
\mbox{}

\mbox{}


% 45.20.-d Formalisms in classical mechanics. 
% 45.20.Jj Lagrangian and Hamiltonian mechanics. 
% 02.30.Hq Ordinary differential equations. 

\mbox{}


\vfill

\noindent $^1$ Telephone: (+34) 91 488 73 91. Fax: (+34) 91 664 74 55. \newline 
\mbox{} \hspace{0.05cm} E-mail: benito.hernandez@urjc.es 

\pagebreak
\begin{flushleft}
{\large {\bf 1. Introduction}}
\end{flushleft}

In terms of coordinates $x_1, \ldots , x_n$, a finite-dimensional Poisson system (for instance, see \cite{olv1} for an introductory review) is a smooth dynamical system defined in a domain $\Omega \subset \mathbb{R}^n$, of the form
\begin{equation}
    \label{nd1nham}
    \dot{x} \equiv \frac{\mbox{d}x}{\mbox{d}t}= {\cal J}(x) \cdot \nabla H(x) 
\end{equation} 
where $x = (x_1, \ldots ,x_n)^T$, superscript $^T$ denotes the transpose matrix, the smooth time-independent first integral $H(x)$ is the Hamiltonian, and the $n \times n$ structure matrix ${\cal J}(x)$ is composed by the structure functions $J_{ij}(x)$ which must be also smooth and verify the Jacobi PDEs: 
\begin{equation}
\label{jacpdes}
     \sum_{l=1}^n 
	\left( \begin{array}{c} J_{li}(x) \partial_l J_{jk}(x) + J_{lj}(x) \partial_l J_{ki}(x) + 
     J_{lk}(x) \partial_l J_{ij}(x) \end{array} \right) = 0 
	\:\; , \;\:\;\: i,j,k=1, \ldots ,n
\end{equation}
where $ \partial_l \equiv \partial / \partial x_l$. Structure functions are also 
skew-symmetric: $J_{ij}(x) =  - J_{ji}(x)$ for all $i,j$.

Poisson systems are present in most fields of nonlinear dynamics, including domains such as mechanics, electromagnetism, optics, population dynamics, control theory, plasma physics, etc. Making use of the Poisson format, many techniques and specific methods adapted to such systems have been developed, including stability analysis, bifurcation methods, characterization of chaotic dynamics, perturbation methods, numerical integration, integability properties, etc. (e.g. see \cite{bn4,bma1} and references therein for the previous statements). Poisson systems constitute a generalization of classical Hamiltonian vector fields \cite{olv1}, allowing for odd-dimensional flows, the classical symplectic matrices of Hamiltonian theory now becoming just special instances of constant structure matrices. Nicely, formal invariance of equations (\ref{nd1nham}) is not limited to canonical transformations, but every smooth diffeomorphism maps a Poisson system into another one. On the other hand, Poisson systems retain the essence of Hamiltonian dynamics: this dynamical equivalence is based on Darboux' theorem. In spite that the Darboux canonical form always exists, Darboux theorem is not constructive (it is an existence theorem) and applies only locally (in the neigborhood of every point, in domains in which the structure matrix has constant rank). However, diverse efforts have been devoted to the explicit and global construction of the Darboux coordinates (for instance, see \cite{bn4}-\cite{cpe1}). This is relevant not only as a natural way to improve the scope of Darboux' theorem, but also as a useful reduction of a Poisson system into a classical Hamiltonian flow (often of lower dimension) with the advantages and plethora of specific methods inherent to the latter format.

In this work, a new procedure for the global construction of the Darboux canonical form is presented. This approach is based on the concept of matrix congruence, and can be applied without the previous determination of the Casimir invariants (it is worth recalling that their prior knowledge is one of the requisites of the standard reduction, thus requiring the integration of the system of PDEs ${\cal{J}}(x) \cdot \nabla C(x)=0$, or solving some equivalent problem \cite{cas1}). Well the opposite, now both the Darboux coordinates and a complete set of independent Casimir invariants arise simultaneously as the outcome of the new reduction algorithm. In addition, the congruence method proceeds only in terms of matrix-algebraic transformations and direct quadratures, thus avoiding the need of previously integrating a system of PDEs. 

The structure of the article is the following. In Section 2 the basic procedure is presented. Its generalization by means of new-time transformations is developed in Section 3. The work concludes in Section 4 with some final remarks. Several examples of physical significance illustrating different aspects of the theory are provided in Sections 2 and 3. 

\mbox{}

\begin{flushleft}
{\large {\bf 2. Global Darboux construction via congruence}}
\end{flushleft}

\begin{flushleft}
\noindent {\bf 2.1. Classical congruence for constant matrices and constant structure matrices}
\end{flushleft}

For the following developments it is necessary to begin by briefly recalling some results about constant matrices (see \cite{ayre} for further details) which shall also be analyzed from the Poisson structure point of view. Every constant skew-symmetric real matrix $A$ of size $n \times n$ and (even) rank $r$ is congruent with matrix ${\cal S}_{(n,r)}$ given by: 
\begin{equation}
\label{snr}
	{\cal S}_{(n,r)}= 
	\left( \begin{array}{cc} 0 & 1 \\ -1 & 0 \end{array} \right) 
	\overbrace{ \oplus \ldots \oplus }^{r/2} 
	\left( \begin{array}{cc} 0 & 1 \\ -1 & 0 \end{array} \right) 
	\oplus O_{(n-r) \times (n-r)}
\end{equation}
where $O_{(n-r) \times (n-r)}$ denotes the $(n-r) \times (n-r)$ null matrix. In other words, there exists a real and invertible $n \times n$ matrix $K$ such that ${\cal S}_{(n,r)} = K \cdot A \cdot K^T$. In fact, such reduction can be accomplished by means of a sequence of elementary row and column transformations. These are performed through left and right matrix multiplications by the corresponding row and column elementary matrices, respectively. Such elementary matrices can be of three types and, in each case, the column elementary transformation matrix is the transpose of its associated row elementary transformation matrix. There are three kinds of elementary transformations: 
\begin{itemize}
\item {\bf Type 1:} Permutation of the $i$-th and $j$-th row (matrix $P_R[i,j]$) or column (matrix $P_C[i,j]$):  
\begin{equation}
\label{eltransf1}
	(P_R[i,j])_{pq} = \left\{ \begin{array}{cl} 
								1 & \mbox{if} \:\; (p,q)=\{(i,j),(j,i),(k,k)\} \:\; 
									\mbox{for} \:\; k \neq i,j \\
								0 & \mbox{otherwise} 
							 \end{array} \right.
\end{equation}
Since $P_R[i,j]$ is symmetric, we now have $P_C[i,j]=(P_R[i,j])^T=P_R[i,j]$. 
\item {\bf Type 2:}  Multiplication by a nonzero constant $c \in \mathbb{R}$ of the $i$-th row ($M_R[i;c]$) or column ($M_C[i;c]$):
\begin{equation}
\label{eltransf2}
	(M_R[i;c])_{jk} = \left\{ \begin{array}{cl} 
								1 & \mbox{if} \:\; j = k \neq i \\
								c & \mbox{if} \:\; j = k = i \\
								0 & \mbox{otherwise} 
							 \end{array} \right.
\end{equation}
As anticipated, it is $M_C[i;c]=(M_R[i;c])^T=M_R[i;c]$.
\item {\bf Type 3:} Linear combination in which $c \in \mathbb{R}$ times the $j$-th row (or column) is added to the 
$i$-th row (or column). These shall be denoted by $L_R[i;c,j]$ and $L_C[i;c,j]$, respectively:
\begin{equation}
\label{eltransf3}
	(L_R[i;c,j])_{kl} = \left\{ \begin{array}{cl} 
								1 & \mbox{if} \:\; k = l \\
								c & \mbox{if} \:\; (k,l) = (i,j) \\
								0 & \mbox{otherwise} 
							 \end{array} \right.
\end{equation}
And $L_C[i;c,j]=(L_R[i;c,j])^T$.
\end{itemize}
As indicated, in congruence transformations the row and column elementary transformations are applied in parallel, the row (column) operations being left (right) multiplications by the row (column) elementary transformation matrices. Since the column transformation matrices are the transpose of the row ones, this produces by construction a congruence transformation of the original skew-symmetric matrix in which the outcome is another skew-symmetric matrix. Thus, the reduction to the canonical form (\ref{snr}) can be always achieved constructively by means of the following algorithm:
\begin{enumerate}
\item Given the skew-symmetric matrix $A \equiv (a_{ij})$, assume without loss of generality $a_{12} \neq 0$. If this is not the case and $A$ is the null matrix, then $A$ is already in canonical form; alternatively if $A$ is not the null matrix, then use of row and column permutations can always be made in order to have a nonzero element in positions (1,2) and (2,1).
\item We now multiply row 1 and column 1 by the nonzero constant $1/a_{12}$. The result is:
\[
	A^* = M_R[1;(1/a_{12})] \cdot A \cdot M_C[1;(1/a_{12})] = 
		\left( \begin{array}{cccc}
			 0                    &  1      & \vline & B_{2 \times (n-2)}     \\
			-1                    &  0      & \vline & \mbox{}                \\ \hline
			\mbox{}               & \mbox{} & \vline & \mbox{}                \\
			-B^T_{(n-2) \times 2} & \mbox{} & \vline & C_{(n-2) \times (n-2)}
		\end{array} \right)
\]
where $B$ and $C$ are submatrices of the sizes indicated by their subindexes. 
\item Using elementary transformations of the third kind $L_R[i;c,j]$ and $L_C[i;c,j]$ the previous matrix $A^*$ can be transformed by congruence into:
\[
	A^{**} = 
		\left( \begin{array}{cccc}
			 0                 &  1      & \vline & O_{2 \times (n-2)}     \\
			-1                 &  0      & \vline & \mbox{}                \\ \hline
			\mbox{}            & \mbox{} & \vline & \mbox{}                \\
			O_{(n-2) \times 2} & \mbox{} & \vline & C^*_{(n-2) \times (n-2)}
		\end{array} \right)
\]
where $O$ denotes the null submatrix and $C^*$ is also a submatrix, both of the given sizes. 
\item Now if $C^*$ is the $(n-2) \times (n-2)$ null submatrix then the reduction is complete. On the contrary, if $C^*$ is not the null submatrix, then the process continues according to steps 1 to 3, this time operating over submatrix $C^*$ until the canonical form (\ref{snr}) is reached.  
\end{enumerate}
According to this reduction procedure, the canonical form is constructed by means of $m$ successive row and column operations ($K_i$ and $K^T_i$ respectively, $i=1, \ldots ,m$) as: 
\[
{\cal S}_{(n,r)}= K_m \cdot K_{m-1} \cdot \ldots \cdot K_1 \cdot A \cdot K_1^T \cdot \ldots \cdot K^T_{m-1} \cdot K^T_m
\] 
Thus, the congruence matrix relating the original matrix $A$ to its canonical form ${\cal S}_{(n,r)}$ is by construction  $K=\prod_{i=1}^mK_{m-i+1}$.

These classical matrix algebra results regarding congruence of skew-symmetric matrices have a meaningful interpretation in the framework of Poisson structures. The starting point is to notice the identity between the set of constant structure matrices for finite-dimensional Poisson systems and the set of constant real skew-symmetric matrices, since the latter are always solutions of the Jacobi PDEs (\ref{jacpdes}). The second issue to be noticed is the identity between the canonical form $(\ref{snr})$ and the Darboux canonical form structure matrix. The third element is the realization of the fact that the transformation rule for structure matrices under diffeomorphisms is a matrix congruence. Recall that after an arbitrary smooth diffeomorphism $y \equiv y(x)$ of a Poisson system (\ref{nd1nham}), Hamiltonian $H(x)$ is transformed in the new Hamiltonian $H^*(y)=H(x(y))$, and the structure matrix ${\cal J}(x)$ becomes a new structure matrix ${\cal J}^* (y)$ according to the rule: 
\begin{equation}
\label{jtransf}
      J^*_{ij}(y) = \sum_{k,l=1}^n \frac{\partial y_i}{\partial x_k} J_{kl}(x) 
	\frac{\partial y_j}{\partial x_l} \;\: , \;\:\;\: i,j = 1, \ldots ,n
\end{equation}
Actually, equation (\ref{jtransf}) can be rewritten as ${\cal J}^*(y) = D(y(x)) \cdot {\cal J}(x) \cdot D^T(y(x))$, where $D(y(x))= \partial (y_1(x), \ldots , y_n(x)) / \partial (x_1, \ldots , x_n)$ is the Jacobian matrix of the diffeomorphic change of coordinates. In the previous case of constant congruence matrices what we actually have is constant Jacobian matrices or, in other words, the Jacobians of invertible linear transformations. Consequently, the congruence reduction just described for constant matrices can be reexpressed as follows: every constant structure matrix can be globally reduced in ${\mathbb R}^n$ to the Darboux canonical form by means of a linear and invertible transformation whose Jacobian matrix is the congruence matrix. Moreover, the reduction can be constructed algorithmically by means of elementary row and column transformation matrices, which themselves implement  elementary invertible linear transformations. Then, when an elementary row transformation (of matrix $K$) and its associated column transformation (of matrix $K^T$) are performed, the constant structure matrix ${\cal J}$ is submitted to the invertible linear transformation $y = K \cdot x$ and the outcome is the transformed structure matrix ${\cal J}^*=K \cdot {\cal J} \cdot K^T$. In this way, successive elementary linear transformations (acting as paired elementary row and column matrix multiplications) produce the overall invertible linear transformation leading to the Darboux canonical form in ${\mathbb R}^n$ for every constant structure matrix. 

This approach is not limited to constant structure matrices, but can be extended to the most general case. This is the purpose of the following two subsections.

\mbox{}

\begin{flushleft}
\noindent {\bf 2.2. Functional congruence for general structure matrices}
\end{flushleft}

Let us now consider the case of a general $n \times n$ structure matrix ${\cal J}(x)$ defined in a domain $\Omega \subset {\mathbb R}^n$ and of constant rank $r$ in $\Omega$. It is thus always possible to construct a congruence (and thus invertible) matrix $K(x)$ such that ${\cal S}_{(n,r)}= K(x) \cdot {\cal J}(x) \cdot K^T(x)$. However, in this framework it is not necessarily true that $K(x)$ is the Jacobian matrix of a diffeomorphism in $\Omega$. The following result puts this issue in perspective: 

\mbox{}

\noindent{\bf Theorem 1.} 
{\em Let ${\cal J}(x)$ be a structure matrix of dimension $n$ and constant rank $r$ in a domain $\Omega \subset {\mathbb R}^n$. Then:
\begin{description}
\item{\mbox{\em a)}} For every $x \in \Omega$ there exists a subdomain $\Omega ^*$, with $x \in \Omega^* \subset \Omega$, such that there exists an $n \times n$ congruence matrix $K(x)$ invertible and smooth everywhere in $\Omega^*$, namely ${\cal S}_{(n,r)}=K(x) \cdot {\cal J}(x) \cdot K^T(x)$ for all $x \in \Omega^*$. 
\item{\mbox{\em b)}} The congruence matrix $K(x)$ considered in item (a) is the Jacobian matrix of a smooth diffeomorphism in that subdomain $\Omega^* \subset \Omega$ (and therefore such diffeomorphism leads to the Darboux canonical form for ${\cal J}(x)$ globally in $\Omega^*$) if and only if the entries $K_{ij}(x)$ of $K(x)$ satisfy everywhere in $\Omega^*$ the conditions:
\begin{equation}
\label{Kjacobiano}
	\frac{\partial K_{ij}(x)}{\partial x_k} = \frac{\partial K_{ik}(x)}{\partial x_j} \;\: , \;\:\;\:\;\: 
	\mbox{for all} \;\:  i,j,k = 1, \ldots ,n.
\end{equation}
\end{description}
}

\mbox{}

\noindent{\bf Proof.} 

Statement (a) can be proved constructively by means of the generalization of the elementary transformations (introduced in the previous subsection) which proceeds as follows. Type 1 transformations remain exactly the same than for the constant matrix case (\ref{eltransf1}), and thus they are still denoted by matrices $P_R[i,j]$ and $P_C[i,j]$. Type 2 transformations now become the multiplication by a smooth function $\xi(x)$ (with $\xi(x) \neq 0$ for every $x \in \Omega^*$) of the $i$-th row (matrix $M_R[i;\xi(x)]$) or column ($M_C[i;\xi(x)]=(M_R[i;\xi(x)])^T$) defined as in the constant case after substitution of the constant $c$ by function $\xi(x)$ in (\ref{eltransf2}). To conclude, Type 3 is given by a linear combination in which $\xi(x)$ times the $j$-th row (or column) is added to the $i$-th row (or column) for $\xi(x)$ an arbitrary smooth function. These shall be now denoted by $L_R[i;\xi(x),j]$ and $L_C[i;\xi(x),j]=(L_R[i;\xi(x),j])^T$, respectively, and again the definition is the same than in the constant case (\ref{eltransf3}) after substitution of the constant $c$ by function $\xi(x)$. In particular, it is worth recalling that every elementary transformation matrix thus defined is by construction smooth and invertible everywhere in a subdomain $\Omega^* \subset \Omega$, but not necessarily everywhere in $\Omega$. The source of this difference are the sets in which the entries of ${\cal J}(x)$ vanish, which may induce a partition of $\Omega$ in subdomains. Regarding smoothness, recall that Poisson systems and structure matrices are defined as smooth. Then, by construction the congruence matrix $K(x)$ is smooth in $\Omega^*$ (see Example 2 for a counterexample proving and illustrating that $\Omega^* \subset \Omega$ may be different from $\Omega$). Now the reduction procedure is entirely similar to the one described for the constant case, the result being a congruence matrix $K(x)=\prod_{i=1}^mK_{m-i+1}(x)$ which is formed by the composition of $m$ elementary transformations of matrices $K_i(x)$, $i=1, \ldots ,m$.  

For part (b), recall that a diffeomorphic transformation leading to the Darboux canonical form must be smooth, and consequently the Jacobian of such transformation must be smooth in $\Omega^*$, which is verified according to statement (a). In addition, matrix $K(x)$ will be the Jacobian matrix of a smooth diffeomorphism $y \equiv y(x)$ provided its entries are of the form $K_{ij}(x)= \partial y_i (x) / \partial x_j$. This is possible \cite{frk1} if and only if:  
\[ 
	\frac{\partial K_{ij}(x)}{\partial x_k} = 
	\frac{\partial ^2 y_i(x)}{\partial x_j \partial x_k} = 
	\frac{\partial K_{ik}(x)}{\partial x_j}
		\;\:\;\: , \;\:\;\: i,j,k = 1, \ldots ,n
\]
These identies are precisely those in (\ref{Kjacobiano}). The proof is thus complete. \hfill $\Box$

\mbox{}

Now some remarks are in order. In first place, it is worth noting that a congruence matrix $K(x)$ verifying the conditions described in item (a) of Theorem 1 is never unique (it is clear that $-K(x)$ is also a congruence matrix if $K(x)$ is, in the same subdomain $\Omega^*$ and having the same properties). Moreover, it is possible that different congruence matrices exist, each defined on different subdomains of $\Omega$ for the same $x \in \Omega$. In general, there are additional degrees of freedom for non-uniqueness to take place, because neither the set of elementary transformations to be applied, nor their order of application, are always unique (this is also true in the case of constant structure matrices, see Example 1). As a result, different congruence matrices may be produced (this issue shall be illustrated in Examples 1 and 2). 

A natural question derived from the point of view of the construction of the Darboux canonical form is whether or not the congruence matrix to be determined is a Jacobian matrix. In this sense, it is interesting to establish a distinction between row and column elementary transformation matrices (ETMs in what follows) defined in the Proof of Theorem 1, and Jacobian elementary transformation matrices (JETMs from now on). A row ETM $K(x)$ and its counterpart the column ETM $K^T(x)$ will be termed JETMs iff $K(x)$ is a Jacobian matrix in the domain of interest. This distinction is relevant because not every ETM is a JETM. The consequences of this are important for the problem of interest, and are developed in the following: 

\mbox{}

\noindent{\bf Theorem 2.} 
{\em Let Let ${\cal J}(x)$ be a structure matrix defined in a domain $\Omega \subset {\mathbb R}^n$, and let $1 \leq i , j \leq n$ be integers with $i \neq j$. Then:
\begin{description}
\item{\mbox{\em a)}} Matrices $P_R[i,j]$ and $P_C[i,j]$ are always JETMs in $\Omega$. The congruence $K(x) \cdot {\cal J}(x) \cdot K^T(x)$, with $K(x)=P_R[i,j]$ produces a transformed structure matrix ${\cal J}^*(y)$ which is the result of performing the diffeomorphic (in fact affine linear) transformation $y = K \cdot x + c$ (with $c$ a vector of arbitrary constants).
\item{\mbox{\em b)}} Matrices $M_R[i;\xi(x)]$ and $M_C[i;\xi(x)]$, with $\xi(x)$ smooth and nonvanishing in $\Omega$, are JETMs in $\Omega$ if and only if $\xi(x) \equiv \xi(x_i)$. The congruence $K(x) \cdot {\cal J}(x) \cdot K^T(x)$, with $K(x)=M_R[i;\xi(x_i)]$ produces a transformed structure matrix ${\cal J}^*(y)$ which is the result of performing the diffeomorphic transformation in $\Omega$ given by $ \{ y_i = \int \xi(x_i) \mbox{\rm d}x_i \; , \;\: y_j = x_j \; , \;\: \mbox{for} \;\: j \neq i \}$. 
\item{\mbox{\em c)}} Matrices $L_R[i;\xi(x),j]$ and $L_C[i;\xi(x),j]$, with $\xi(x)$ smooth in $\Omega$, are JETMs in $\Omega$ if and only if $\xi(x) \equiv \xi(x_j)$. The congruence $K(x) \cdot {\cal J}(x) \cdot K^T(x)$, with $K(x)=L_R[i;\xi(x_j),j]$ produces a transformed structure matrix ${\cal J}^*(y)$ which is the result of performing the diffeomorphic transformation in $\Omega$ given by $\{ y_i = x_i + \int \xi(x_j) \mbox{\rm d}x_j \; , \;\: y_k = x_k \; , \;\: \mbox{for} \;\: j \neq i \;\: \mbox{and} \;\: k \neq i \}$.
\item{\mbox{\em d)}} A sufficient condition for a matrix to be a Jacobian in $\Omega$ is that it is the product of JETMs in $\Omega$. This condition is not necessary, namely it is possible that a Jacobian matrix in $\Omega$ is the product of ETMs in $\Omega$ such that not all of them are JETMs.
\end{description}
}

\mbox{}

\noindent{\bf Proof.} 

For part (a), taking the set of linear transformations $y=P_R[i,j] \cdot x + c$, with $c$ a vector of arbitrary constants, we have that the constant matrix $P_R[i,j]$ is the Jacobian of such transformations. According to rule (\ref{jtransf}) the congruence transformation ${\cal J}^*(y) = P_R[i,j] \cdot {\cal J}(x) \cdot (P_R[i,j])^T$ leads to the structure matrix ${\cal J}^*(y)$ in the new variables.  

Regarding item (b), both $M_R[i;\xi(x)]$ and $M_C[i;\xi(x)]=(M_R[i;\xi(x)])^T$ are the same diagonal matrix. In this case, the only nontrivial identities (\ref{Kjacobiano}) are those of the form $\partial _j K_{ii} = \partial _j \xi(x) = \partial _i K_{ij}$, for $i \neq j$, where it is necessarily $\partial _i K_{ij} =0$. The identities then amount to $\partial _j \xi(x) = 0$ for all $j \neq i$, or $\xi(x) \equiv \xi(x_i)$. Now if we compute the Jacobian of transformation $\{ y_i = \int \xi(x_i) \mbox{\rm d}x_i \; , \;\: y_j = x_j \; , \;\: \mbox{for} \;\: j \neq i \}$ it can be seen that it is precisely $M_R[i;\xi(x_i)]$. To conclude, use of rule (\ref{jtransf}) completes the proof of this statement. 

For part (c), consider first matrix $K(x)=L_R[i;\xi(x),j]$. Recall that the only nonzero entries of the $i$-th row of such matrix are that of value $1$ at position $(i,i)$ and that of value $\xi(x)$ at position $(i,j)$, with $j \neq i$. Accordingly, the only nontrivial PDE from the set (\ref{Kjacobiano}) is given by $\partial _k K_{ij} = \partial _j K_{ik}$, with $k \neq j$ (note that case $k=j$ amounts to a trivial identity). This leads to $\partial _k \xi(x) = 0$ for $k \neq j$, or $\xi(x) \equiv \xi(x_j)$. The reasoning is entirely similar for matrix $L_C[i;\xi(x),j]$, where we also conclude $\xi(x) \equiv \xi(x_j)$, consistently with the fact that $L_C[i;\xi(x_j),j]=(L_R[i;\xi(x_j),j])^T$. Accordingly, the Jacobian of the transformation family $\{ y_i = x_i + \int \xi(x_j) \mbox{\rm d}x_j \; , \;\: y_k = x_k \; , \;\: \mbox{for} \;\: j \neq i \;\: \mbox{and} \;\: k \neq i \}$ is precisely $L_R[i;\xi(x_j),j]$. Then rule (\ref{jtransf}) accounts for the congruence transforming ${\cal J}(x)$ into ${\cal J}^*(y)$ with $L_R[i;\xi(x_j),j]$ as congruence matrix. 

Regarding statement (d), in one sense it is known that the matrix product of Jacobian matrices produces the Jacobian matrix of the composite transformation. In the opposite sense, it can be shown by means of a counterexample that the product of ETMs not all of them being JETMs can produce a Jacobian matrix. Such counterexample shall be provided in Example 7, to which the reader is referred. This completes the proof of Theorem 2. \hfill $\Box$

\mbox{}

According to the previous results, the procedure to be used in order to construct globally the Darboux canonical form consists of the use of JETMs in order to find a congruence between the initial structure matrix ${\cal J}(x)$ and the Darboux canonical matrix ${\cal S}_{(n,r)}$. The sucessive composition of smooth diffeomorphic transformations leads to a direct smooth diffeomorphism $y \equiv y(x)$ reducing ${\cal J}(x)$ to the Darboux form. In this sense, some remarks deserve an explicit mention:

\begin{itemize} 
\item The first one is that the method is constructive and proceeds entirely in terms of matrix algebra: the integration of ODEs or PDEs is not necessary in order to find the Jacobian of the transformation. 
\item In addition, the explicit diffeomorphic transformation $y \equiv y(x)$ can be found from the Jacobian $K(x) \equiv (K_{ij}(x))$ by means of simple quadratures of the form $y_i(x)= \int K_{ij}(x) \mbox{\rm d}x_j = \int (\partial y_i(x) / \partial x_j) \mbox{\rm d}x_j \; , \:\; j=1, \ldots ,n$. The compatible solutions of these quadratures provide the diffeomorphism, up to an arbitrary set of integration constants. (See Examples 1, 3, 4, 5, 7, and 8 for instances of this procedure). 
\item Alternatively, the explicit diffeomorphic transformation $y \equiv y(x)$ can also be found as the composition of the successive elementary transformations associated with each JETM performed, which were explicitly characterized in Theorem 2 (see Example 5 for an illustration).
\item Contrarily to the case in other reduction procedures, now a complete set of functionally independent Casimir invariants $\{ C_{r+1}(x), \ldots ,C_n(x) \}$ of ${\cal J}(x)$ needs not to be known in advance. Recall that such previous knowledge implies the resolution of the set of coupled PDEs ${\cal J}(x) \cdot \nabla C(x) = 0$ (or some equivalent formulation of the problem \cite{cas1}). In fact, the use of the congruence method provides the Casimir invariants through the purely algebraic construction of matrix $K(x)$: notice that the congruence between ${\cal J}(x)$ and ${\cal S}_{(n,r)}$ as defined in (\ref{snr}) implies that a complete set of functionally independent Casimir invariants of ${\cal J}(x)$ is given by the expressions: $C_i(x) = y_i(x) \; , \:\; i=r+1, \ldots ,n$, obtained from the last $(n-r)$ equations of transformation $y \equiv y(x)$. See Examples 3, 4, 5 and 7 for instances of this issue. 
\item When it exists, a Darboux congruence reduction is global in scope (in the sense of being defined in a domain, and not just in a point-dependent neighborhood). Recall that Darboux' theorem is not constructive (it is an existence theorem) and it is only local in scope, being valid in principle in a neighborhood of every point $x$. In practice, such reduction has been constructed globally for many solution families of structure matrices \cite{bn4}-\cite{cpe1}, while in many other cases there is no known global reduction (which, in fact, might not exist). In a similar way, the existence of a congruence reduction is not guaranteed in advance (in other words, a congruence matrix always exists but it is not necessarily a Jacobian matrix, as shown in Theorem 1). Examples 1 and 2 provide instances in this sense.  
\end{itemize} 

All features mentioned throughout this subsection (including the counterexamples proving some of the statements made) shall be illustrated in detail. 

\mbox{}

\begin{flushleft}
\noindent {\bf 2.3. Examples}
\end{flushleft}

\noindent {\bf Example 1.} {\em Two-dimensional Poisson systems.}

\mbox{}

It is well-known that every 2-d smooth skew-symmetric matrix is a structure matrix: 
\begin{equation}
\label{j2dgeneral}
	{\cal J}(x_1,x_2) = \left( \begin{array}{cc} 0 & f(x_1,x_2) \\ - f(x_1,x_2) & 0 \end{array} \right)
\end{equation}
Provided $f(x_1,x_2) \neq 0$ in $\Omega \subset {\mathbb R}^2$, we have Rank(${\cal J}$) $=2$ in $\Omega$. 
The congruence matrix is not unique. One possibility is $K(x) = M_R(1;1/f(x_1,x_2)) = \mbox{diag} (1/f(x_1,x_2),1 )$. In this case, application of conditions (\ref{Kjacobiano}) shows that $K(x)$ is a Jacobian matrix iff $\partial _2 f(x_1,x_2) =0$, namely $f(x_1,x_2) \equiv f(x_1)$. Then the smooth transformation leading to the Darboux canonical form is $\{ y_1 = \int (1/f(x_1)) \mbox{d} x_1, y_2=x_2 \}$. 
Alternatively, it is possible a different choice of the type $K^*(x) = M_R(2;1/ f(x_1,x_2)) = \mbox{diag} 
( 1,1/f(x_1,x_2) )$. In this case, $K^*(x)$ is a Jacobian matrix iff $\partial _1 f(x_1,x_2) =0$, or $f(x_1,x_2) \equiv f(x_2)$, with transformation $\{ y_1=x_1, y_2 = \int (1/f(x_2)) \mbox{d}x_2 \}$. In addition, when $f(x_1,x_2) = c \neq 0$ is constant, this shows non-uniqueness of the congruence matrix for constant skew-symmetric matrices, since we obtain $K = \mbox{diag} (1/c,1)$ and $K^* = \mbox{diag} (1,1/c)$ which are equal only in the trivial case $c=1$. On the other hand, when $f$ is a simultaneous function of $x_1$ and $x_2$ none of the congruence matrices is a Jacobian matrix, and the Darboux reduction method is not applicable. A generalization of the congruence reduction procedure applicable to such case will be developed in the next section (see also Example 6). 

\mbox{}

\noindent {\bf Example 2.} {\em Lie-Poisson $so(3)$ structure matrix.}

\mbox{}

We shall now consider the $so(3)$ structure matrix, well known in mechanics \cite{olv1}:
\begin{equation}
\label{jso3}
	{\cal J}(x) = \left( \begin{array}{ccc} 0 & -x_3 & x_2 \\ x_3 & 0 & -x_1 \\ -x_2 & x_1 & 0 \end{array} \right)
\end{equation}
Since ${\cal J}$ is smooth and Rank(${\cal J}$) $=2$ everywhere except at the origin, we now have $\Omega = {\mathbb R}^3 - \{(0,0,0)\}$. Following the reduction algorithm to form (\ref{snr}), we can define the following ETMs: 
\[
	K_1(x)= M_R[1;-1/x_3]=(M_C[1;-1/x_3])^T = 
\left( \begin{array}{ccc} -1/x_3 & 0 & 0 \\ 0 & 1 & 0 \\ 0 & 0 & 1 \end{array} \right) 
\]
\[
	K_2(x)= L_R[3;x_2/x_3,2]=(L_C[3;x_2/x_3,2])^T = 
\left( \begin{array}{ccc} 1 & 0 & 0 \\ 0 & 1 & 0 \\ 0 & x_2/x_3 & 1 \end{array} \right) 
\]
\[
	K_3(x)= L_R[3;-x_1,1]=(L_C[3;-x_1,1])^T = 
\left( \begin{array}{ccc} 1 & 0 & 0 \\ 0 & 1 & 0 \\ -x_1 & 0 & 1 \end{array} \right) 
\]
We then have the following congruence reduction: 
\[
	K_3(x) \cdot K_2(x) \cdot K_1(x) \cdot {\cal J}(x) \cdot K_1^T(x) \cdot K_2^T(x) \cdot K_3^T(x) = {\cal S}_{(3,2)} =
	\left( \begin{array}{ccc} 0 & 1 & 0 \\ -1 & 0 & 0 \\ 0 & 0 & 0 \end{array} \right) \;\: , \;\:\;\: 
\]
Now it is relevant to note that $K_1(x)$ and $K_2(x)$ are smooth and invertible in $\Omega^*_+ \subset \Omega$ and $\Omega^*_- \subset \Omega$, where $\Omega^*_+ = \{x \in \mathbb{R}^3: x_3>0 \}$, and $\Omega^*_- = \{x \in \mathbb{R}^3: x_3<0 \}$.
This is the counterexample anticipated in the proof of Theorem 1: we see thus that for every $x \in \Omega$ there exists a subdomain $\Omega^* \subset \Omega$ in which congruence is possible in terms of elementary matrices smooth and invertible in $\Omega^*$ (in this case, note that if we are interested in congruence for points different from the origin but lying on the plane $x_3 =0$, then there are analogous constructions in which the singular planes are $x_1=0$ or $x_2=0$, the congruence matrices being smooth at the points for which $x_3=0$). 

Notice in addition that matrices $K_1$ and $K_2$ are not JETMs, while $K_3$ is. The global congruence matrix is: 
\[
	K(x) = K_3(x) \cdot K_2(x) \cdot K_1(x) = 
	\left( \begin{array}{ccc} -1/x_3 & 0 & 0 \\ 0 & 1 & 0 \\ x_1/x_3 & x_2/x_3 & 1 \end{array} \right) \;\: , \;\:\;\: 
\]
It is simple to verify that $K(x)$ is not a Jacobian matrix, for instance it is $\partial_3 K_{11}(x) \neq \partial_1 K_{13}(x)$. The next section develops an extension of the congruence method leading to the Darboux reduction for matrices such as this one (see Example 7). 

It is worth giving also an illustration of the remarks presented after Theorem 1, in which non-uniqueness of congruence matrices was anticipated. For instance, the following is an alternative sequence: $K^*_1 = M_R[3;x_3/x_2]$, $K^*_2=L_R[3;1,2]$, $K_3^*=L_R[3;x_1/x_2,1]$, and $K_4^* = M_R[1;1/x_3]$. The outcome is another congruence reduction to ${\cal S}_{(3,2)}$ with a different congruence matrix $K^* = \prod _{i=1}^4 K^*_{5-i}$ which is not a Jacobian (in addition, note that $K_i^*$ is not a JETM for $i=1,3,4$).  

\mbox{}

\pagebreak
\noindent {\bf Example 3.} {\em Kermack-McKendrick system.}

\mbox{}

The following structure matrix appears in the Poisson formulation of the well-known Kermack-McKendrick model \cite{kmck}: 
\begin{equation}
\label{jkmck}
	{\cal J}(x) = bx_1x_2 \left( \begin{array}{ccc} 0 & 1 & -1 \\ -1 & 0 & 1 \\ 1 & -1 & 0 \end{array} \right)  
\end{equation}
Matrix (\ref{jkmck}) is defined for $b>0$ and $\Omega = \mbox{\rm Int} (\mathbb{R}_+^{3})$ and has been classified as a multiseparable structure matrix (see \cite{bn4} for details) where the global reduction to the Darboux canonical form was constructed. However, such reduction is quite elaborate, and a much simpler procedure is now possible in terms of the congruence approach. This is done by literally following the steps of the algorithm in terms of four JETMs: $K_1(x) = L_R[3;1,1]$, $K_2(x) = L_R[3;1,2]$, $K_3(x) = M_R[1;1/x_1]$, $K_4(x) = M_R[2;1/(bx_2)]$. Since all of them are JETMs, the outcome is an overall Jacobian congruence matrix: 
\[
	K(x) = K_4(x) \cdot K_3(x) \cdot K_2(x) \cdot K_1(x) = 
	\left( \begin{array}{ccc} 1/x_1 & 0 & 0 \\ 0 & 1/(bx_2) & 0 \\ 1 & 1 & 1 \end{array} \right)  
\]
In other words, $K(x) \cdot {\cal J}(x) \cdot K^T(x) = {\cal S}_{(3,2)}$. In addition, the congruence matrix $K(x)$ leads to the direct construction of the diffeomorphism $\{y_1(x) = \log x_1 ; \: y_2(x) = (1/b) \log x_2; \: y_3 (x) = x_1+x_2+x_3\}$, as well as to the only independent Casimir invariant of ${\cal J}(x)$, which is $C(x)=y_3(x) = x_1+x_2+x_3$.

\mbox{}

\noindent {\bf Example 4.} {\em Poisson bracket for the Toda lattice.}

\mbox{}

Toda system, when expressed in Flaschka's variables \cite{dam1,dam2} $(\alpha_1, \ldots , \alpha_{N-1}, \beta_1, \ldots , \beta_N)$ has a Poisson structure of rank $2(N-1)$ with brackets $\{ \alpha_i , \beta_i \} = - \alpha_i$, $\{ \alpha_i , \beta_{i+1} \} = \alpha_i$, while the rest of elementary brackets vanish. For the sake of conciseness the case $N=3$ shall be considered in what follows (the procedure is analogous for arbitrary $N$, but $N=3$ is the first instance of rank $ \geq 4$, a case in which the global Darboux reduction is not known in general \cite{bma1}). 

Let $( \alpha_1, \alpha_2, \beta_1, \beta_2, \beta_3 ) \equiv (x_1, \ldots , x_5)$. The reduction is performed by means of five consecutive JETMs: $K_1(x)=L_R[4;1,3]$, $K_2(x)=L_R[5;1,4]$, $K_3(x)=M_R[1;-1/x_1]$, $K_4(x)=M_R(2;-1/x_2)$, $K_5(x)=P_R[2,3]$. The total Jacobian congruence matrix thus obtained is: 
\begin{equation}
\label{todajacobian}
	K(x) = K_5(x) \cdot K_4(x) \cdot K_3(x) \cdot K_2(x) \cdot K_1(x) = 
		\left(  \begin{array}{ccccc} 
		-1/x_1 & 0 & 0 & 0 & 0 \\ 0 & 0 & 1 & 0 & 0 \\ 0 & -1/x_2 & 0 & 0 & 0 \\ 
		0 & 0 & 1 & 1 & 0 \\ 0 & 0 & 1 & 1 & 1 
		\end{array} \right)
\end{equation}
It is then $K(x) \cdot {\cal J}(x) \cdot K^T(x) = {\cal S}_{(5,4)}$. From (\ref{todajacobian}) it is direct to find the Darboux coordinates $\{ y_1(x) = - \log x_1 ; \; y_2(x) = x_3 ; \; y_3(x) = - \log x_2 ; \; y_4(x) = x_3 + x_4 ; \; y_5(x) = x_3 + x_4 + x_5 \}$, and the only independent Casimir invariant is just $C(x) = y_5(x) = x_3 + x_4 + x_5$.

\mbox{}

\pagebreak
\noindent {\bf Example 5.} {\em Separable structure matrices.}

\mbox{}

Separable structure matrices \cite{sep1} are defined in terms of a set of smooth nonvanishing one-variable functions $\varphi_i(x_i)$, $i=1, \ldots ,n$, and a constant $n \times n$ skew-symmetric matrix $A=(a_{ij})$ of rank $r$. Thus, the entries of a separable structure matrix ${\cal J}(x)$ are of the form $J_{ij}(x)=a_{ij} \varphi_i(x_i) \varphi_j(x_j)$, for all $i,j=1, \ldots ,n$. As shown, separable structure matrices involve many different Poisson systems of physical interest (see \cite{sep1} for details). Here we shall see how both the construction of the Darboux canonical form and the determination of Casimir invariants arise naturally in this case. For this, we consider first the Jacobian congruence matrix $K_1(x) = \mbox{\rm diag}(1/ \varphi_1(x_1), \ldots , 1/ \varphi_n(x_n))$. This matrix has an obvious decomposition in terms of $n$ JETMs $K^{[1]}, \ldots ,K^{[n]}$ of the form $\left( K^{[i]} \right)_{jk} = \delta_{jk} [1+(-1+1/ \varphi_i(x_i))\delta_{ki}]$, for $i,j,k= 1, \ldots , n$, where symbol $\delta$ denotes Kronecker's delta. The outcome is $K_1(x) \cdot {\cal J}(x) \cdot K_1^T(x) = A$. According to the theory reviewed in Subsection 2.1, there exists a second (constant, and then Jacobian) congruence matrix $K_2 = P \equiv (p_{ij})$ such that $P \cdot A \cdot P^T = {\cal S}_{(n,r)}$. As it is known from the congruence algorithm for constant matrices, $K_2$ also has a decomposition in terms of constant JETMs. 
The joint Jacobian $K(x)=K_2 \cdot K_1(x)$ has entries $K_{ij}(x) = p_{ij}/ \varphi_j(x_j)$, for $i,j= 1, \ldots , n$. After straightforward integration, this provides the diffeomorphism leading to the Darboux canonical form as: 
\begin{equation}
\label{septransf}
	y_i(x) = \sum_{j=1}^n p_{ij} \int \frac{1}{\varphi_j(x_j)} \mbox{\rm d}x_j \:\; , \:\;\:\; i= 1, \ldots , n
\end{equation}
In addition, the procedure constructs also for ${\cal J}(x)$ a complete set of independent Casimir invariants which does not need to be known in advance, given by $\{ C_{r+1}(x) = y_{r+1}(x), \ldots ,C_n(x)=y_n(x) \}$ in (\ref{septransf}). This approach is equivalent to, but clearly more efficient than, the one in \cite{sep1}.  

\mbox{}

This concludes the present examples. We can now proceed to the analysis of the problem in its full generality. This is the aim of the next section. 

\mbox{}

\begin{flushleft}
{\large {\bf 3. Generalization by means of new-time transformations}}
\end{flushleft}

\begin{flushleft}
\noindent {\bf 3.1. Generalization of the method}
\end{flushleft}

A new-time transformation (NTT in what follows) of a Poisson system (\ref{nd1nham}) is defined as a time reparametrization of the form:
\begin{equation}
\label{ntt0}
	\mbox{\rm d}\tau = \frac{1}{\eta (x)} \mbox{\rm d}t
\end{equation}
where $t$ is the original time variable, $\tau$ is the new time and $\eta (x) : \Omega \subset 
\mathbb{R}^n \rightarrow \mathbb{R} - \{ 0 \}$ is a function which does not vanish and is smooth in $\Omega$. Accordingly, after NTT 
(\ref{ntt0}) an arbitrary Poisson system (\ref{nd1nham}) defined in $\Omega$ becomes: 
\[
	\frac{\mbox{\rm d}x}{\mbox{\rm d} \tau} = \eta (x){\cal J}(x) \cdot \nabla H (x)
\]
NTTs are necessary for the construction of Poisson structures in different physical problems \cite{bma1},\cite{mm1}-\cite{agfs}. Moreover, in the particular (but prominent) case of classical Hamiltonian systems, the use of NTTs is well-known for the integrability analysis \cite{hgdr}-\cite{gor1}, in stability theory \cite{mha}, etc. Given that NTTs preserve topology in phase-space, very often the global reduction to the Darboux canonical form is possible by means of a diffeomorphism followed by an NTT \cite{bma1},\cite{bn0}-\cite{pd14}. Note however that NTTs do not preserve in general the Poisson structure, namely $\eta(x) {\cal J}(x)$ may not be a structure matrix in spite that ${\cal J}(x)$ is. In previous articles \cite{bma1,ntt2} the following results were developed: 

\mbox{}

{\em Let ${\cal J}(x)$ be an $n \times n$ structure matrix of constant rank $r$ defined in a domain $\Omega \subset \mathbb{R}^n$, let $C(x)$ be a Casimir invariant of ${\cal J}(x)$ in $\Omega$, and let $\eta (x)$ be a smooth function in $\Omega$. Then: 
\begin{itemize}
\item Property 1: The product $\eta (x) {\cal J}(x)$ is a structure matrix in $\Omega$ for every smooth function $\eta (x)$ if and only if $r \leq 2$.
\item Property 2: $C(x) {\cal J}(x)$ is a structure matrix everywhere in $\Omega$ for arbitrary $r$.
\item Property 3: For $n \geq 4$, if ${\cal J}(x)$ is symplectic (namely $r=n$) then the product $\eta (x) {\cal J}(x)$ is a structure matrix in $\Omega$ if and only if $\eta (x)$ is a constant. 
\end{itemize}
}
If, according to the previous three properties, the product $\eta(x) {\cal J}(x)$ remains a structure matrix after NTT (\ref{ntt0}), then function $\eta (x)$ is termed a reparametrization factor for ${\cal J}(x)$. This framework can be employed in order to generalize the congruence procedure for the global construction of the Darboux canonical form:  

\mbox{}

\noindent{\bf Theorem 3.} 
{\em Let ${\cal J}(x)$ be a structure matrix of constant rank $r$ defined in a domain $\Omega \subset {\mathbb R}^n$, and let $\eta(x): \Omega \subset \mathbb{R}^n \rightarrow \mathbb{R} - \{ 0 \}$ be a function smooth in $\Omega$. Then the Darboux canonical form can be constructed globally in $\Omega$ for ${\cal J}(x)$ provided there exists in $\Omega$ a smooth Jacobian congruence matrix $K(x)$ such that:
\begin{equation}
\label{genercongr}
	K(x) \cdot {\cal J}(x) \cdot K^T(x) = \frac{1}{\eta(x)} {\cal S}_{(n,r)}
\end{equation}
}

\mbox{}

\noindent{\bf Proof.} 

The proof is constructive. After a diffeomorphism $y \equiv y(x)$ of Jacobian $K(x)$, matrix ${\cal J}(x)$ becomes the structure matrix ${\cal J}^*(y) = (1 / \eta(x(y)) {\cal S}_{(n,r)}$. Since both ${\cal S}_{(n,r)}$ and ${\cal J}^*(y)$ are structure matrices, we conclude that $\eta (x(y))$ is by definition a reparametrization factor of ${\cal J}^*(y)$. Consequently, a smooth diffeomorphism $y \equiv y(x)$ of Jacobian $K(x)$ applied to the initial Poisson system of structure matrix ${\cal J}(x)$, followed by an NTT of the form $\mbox{\rm d}\tau = (1/ \eta (x(y)))\mbox{\rm d}t$ performed on the transformed Poisson system of structure matrix ${\cal J}^*(y)$ lead to the construction of the global Darboux reduction for ${\cal J}(x)$ in $\Omega$. \hfill $\Box$

\mbox{} 

Equation (\ref{genercongr}) and the method just described constitute a generalization of the congruence method as developed in Section 2, since now there exists an additional degree of freedom (associated with the reparametrization factor) allowing the Darboux reduction. In other words, it can be said that the congruence method shown in Section 2 corresponds to the particular case $\eta (x) = 1$ in identity (\ref{genercongr}). As we shall see in Subsection 3.2 by means of some examples, the use of NTTs increases significantly the scope of the method.  

\mbox{}

\begin{flushleft}
\noindent {\bf 3.2. Examples}
\end{flushleft}

\noindent {\bf Example 6.} {\em Two-dimensional Poisson systems (continued).}

\mbox{}

As shown in Example 1, two-dimensional structure matrices (\ref{j2dgeneral}) cannot be reduced to Darboux canonical form by means of a pure congruence, except in some special cases. However, the reduction is now direct in general provided that the rank is 2 in the domain $\Omega$ of interest. This implies $f(x_1,x_2) \neq 0$ everywhere in $\Omega$, and allows the direct Darboux reduction by means of the time reparametrization $\mbox{d} \tau = f(x_1,x_2) \mbox{d}t$, which in the present framework appears as a special case of the method just exposed in Subsection 3.1. 

\mbox{}

\noindent {\bf Example 7.} {\em Lie-Poisson $so(3)$ structure matrix (continued).}

\mbox{}

Consider the structure matrix (\ref{jso3}), on which the following six JETMs are performed: $K_1(x)=M_R[1;x_1]$, $K_2(x)=M_R[2;x_2]$, $K_3(x)=M_R[3;x_3]$, $K_4(x)=L_R[1;1,2]$, $K_5(x)=L_R[3;1,1]$, $K_6(x)=M_R[1;-1]$. The composite congruence matrix is the Jacobian: 
\begin{equation}
\label{jacex7}
	K(x) = \prod_{i=1}^6 K_{7-i} (x) = 
	\left( \begin{array}{ccc} -x_1 & -x_2 & 0 \\ 0 & x_2 & 0 \\ x_1 & x_2 & x_3 \end{array} \right) 	
\end{equation}
This leads to the transformed structure matrix ${\cal J}^*(y) = K(x) \cdot {\cal J}(x) \cdot K^T(x) = x_1(y)x_2(y)x_3(y) {\cal S}_{(3,2)}$. From Jacobian (\ref{jacex7}) it is direct to integrate the transformation $\{y_1(x) = -(x_1^2+x_2^2)/2 ; \: y_2(x) = x_2^2/2; \: y_3 (x) = (x_1^2+x_2^2+x_3^2)/2 \}$, as well as the only independent Casimir invariant $C(x)=y_3(x) = (x_1^2+x_2^2+x_3^2)/2$ of ${\cal J}(x)$. Now assume without loss of generality that $x \in \mbox{\rm Int} (\mathbb{R}_+^{3})$ (similar procedures can be determined for any other case). Then $\eta(y) = x_1(y)x_2(y)x_3(y) = 2 \sqrt{2} [y_2(-y_1-y_2)(y_1+y_3)]^{1/2}$ and the NTT $\mbox{\rm d} \tau = \eta (y) \mbox{\rm d}t$ completes the construction of the Darboux canonical form. 

In order to provide additional perspective, let us consider again structure matrix (\ref{jso3}), this time with the following ETMs:
$\tilde{K}_1(x)= L_R[3;x_1/x_3,1]$, $\tilde{K}_2(x)= L_R[3;x_2/x_3,2]$, and $\tilde{K}_3(x)= M_R[3;x_3]$. Note that $\tilde{K}_1(x)$ and $\tilde{K}_2(x)$ are not JETMs, while $\tilde{K}_3(x)$ is. These tree ETMs produce the congruence matrix:
\[
	\tilde{K}(x) = \tilde{K}_3(x) \cdot \tilde{K}_2(x) \cdot \tilde{K}_1(x) = 
	\left( \begin{array}{ccc} 1 & 0 & 0 \\ 0 & 1 & 0 \\ x_1 & x_2 & x_3 \end{array} \right) 	
\]
We then have $\tilde{K}(x) \cdot {\cal J}(x) \cdot \tilde{K}^T(x) = (-x_3) {\cal S}_{(3,2)}$. Moreover, $\tilde{K}(x)$ is a Jacobian matrix, which can be easily integrated to provide the transformation $\{ y_1(x) = x_1, y_2(x) = x_2, y_3(x) = (x_1^2+x_2^2+x_3^2)/2 \}$. As a consequence, this implies that a Casimir function for ${\cal J}(x)$ in (\ref{jso3}) is $C(x) = y_3(x) = (x_1^2+x_2^2+x_3^2)/2$. In addition, this counterexample proves that a Jacobian matrix may arise as the product of ETMs, not all of them being JETMs, as anticipated in the proof of Theorem 2(d). Thus, the reduction to Darboux canonical form is completed by means of the time reparametrization $\mbox{d} \tau = - x_3(y) \mbox{d}t = \mp (2y_3-y_1^2-y_2^2)^{1/2} \mbox{d}t$.

\mbox{}

\noindent {\bf Example 8.} {\em D$_\psi$-solutions.}

\mbox{}

Let us now look at some instances of D$_\psi$-solutions (a subset of D-solution structure matrices, see \cite{dist} for details). In first place, consider the following structure matrix, previously analyzed in \cite{dist}: 
\begin{equation}
\label{dpsi}
	{\cal J}(x) = \left( \begin{array}{cccc}
	 0 & \psi (C_3,C_4) & - \psi (C_3,C_4) & - \psi (C_3,C_4) \\
	- \psi (C_3,C_4)  &  0 & 0 & \psi (C_3,C_4) \\ 
	\psi (C_3,C_4) & 0 & 0 & - \psi (C_3,C_4)   \\ 
	\psi (C_3,C_4) & - \psi (C_3,C_4) & \psi (C_3,C_4) & 0 \\
	\end{array} \right) 
\end{equation}
In (\ref{dpsi}), $\psi (z_1,z_2)$ is a smooth nonvanishing function of two real variables, while $C_3(x)=x_2+x_3$ and $C_4(x)=x_1+x_2+x_4$ are two independent Casimir invariants (since Rank(${\cal J}$)$=2$). We proceed to the reduction to the Darboux canonical form by first using three JETMs: $K_1(x)=L_R(4;1,1)$, $K_2(x)=L_R(4;1,2)$ and $K_3(x)=L_R(3;1,2)$. The outcome is: 
\[
K_3 \cdot K_2 \cdot K_1 \cdot {\cal J}(x) \cdot K_1^T \cdot K_2^T \cdot K_3^T = \psi (C_3(x),C_4(x)) {\cal S}_{(4,2)} = \psi(y_3,y_4) {\cal S}_{(4,2)}
\]
As usual, we denote as $y \equiv y(x)$ the new variables $\{y_1(x) = x_1 ; \: y_2(x) = x_2; \: y_3(x) = x_2+x_3; \: y_4(x) = x_1+x_2+x_4 \}$ obtained from the linear diffeomorphism of constant Jacobian $K = K_3 \cdot K_2 \cdot K_1$. Now the reduction is completed with NTT $\mbox{\rm d}\tau = \psi(y_3,y_4) \mbox{\rm d}t$.

Actually this example can be generalized to arbitrary dimension and rank as follows: consider an $n \times n$ skew-symmetric constant real matrix $A$ of rank $r$, and let $\{ v_{r+1}, \ldots ,v_n \}$ be a basis of Ker$\{ A \}$. In addition, let $\psi (z_1, \ldots ,z_{n-r})$ be a smooth nonvanishing function. Then matrix ${\cal J}(x)=\psi (v_{r+1} \cdot x, \ldots , v_n \cdot x) A$ is a particular case of D$_\psi$-solution of dimension $n$ and rank $r$ which generalizes (\ref{dpsi}). In addition, reduction to the Darboux canonical form can be carried out by a procedure entirely analogous to the one just described for (\ref{dpsi}). In fact, notice that (according to the theory presented in Subsection 2.1) there exists a constant (thus Jacobian of a linear transformation) matrix $K$ such that $K \cdot A \cdot K^T = {\cal S}_{(n,r)}$, thus the outcome of such diffeomorphism being ${\cal J}^*(y)=\psi ^* (y_{r+1}, \ldots ,y_n) {\cal S}_{(n,r)}$ in the new variables $y \equiv y(x)$, where $\psi ^*$ is also smooth and nonvanishing (not necessarily equal to $\psi$). The procedure is completed with NTT $\mbox{\rm d}\tau = \psi ^* (y_{r+1}, \ldots ,y_n) \mbox{\rm d}t$. It is worth noting that in \cite{dist} the global Darboux reduction for non-constant D-solutions was only considered for the case $n=3$, $r=2$. Accordingly, this is the first time in the literature (to the author's knowledge) that such construction is performed for a family of non-constant D-solutions of arbitrary dimension and rank. 

\mbox{}

\begin{flushleft}
{\large {\bf 4. Final remarks}}
\end{flushleft}

In this contribution, an algebraic method for the global construction of the Darboux canonical form has been developed. Such procedure has several remarkable features: (i) it is applicable to structure matrices of arbitrary dimension, rank and degree of nonlinearity; (ii) it leads to the simultaneous determination of a complete set of independent Casimir invariants; (iii) it does not require the integration of ODEs or PDEs, but it proceeds entirely in terms of elementary algebraic matrix operations and direct quadratures; (iv) it is inclusive, in the sense that every structure matrix globally reducible to the Darboux canonical form has by definition a functional congruence matrix globally defined providing such reduction, namely the Jacobian of the transformation. (Similarly, in the generalized case of diffeomorphism followed by an NTT, the procedure is also inclusive, in the same sense just described); and (v) reciprocally, if a structure matrix has no global diffeomorphic reduction to Darboux canonical form, then a Jacobian congruence matrix globally defined does not exist and the congruence procedure is not applicable. Consequently, if we consider the standard approach consisting of a previous characterization of the Casimir invariants followed by a reduction to the level set of the symplectic leaves, then we can conclude that the congruence procedure provides new insight as well as some improvements that may lead to further developments in the future.  

\mbox{}

\mbox{}

\noindent {\bf Acknowledgements}

\noindent The author acknowledges the Ministerio de Econom\'{\i}a y Competitividad of Spain for Projects 
MTM2014-53703-P and MTM2016-80276-P, as well as financial support from Universidad Rey Juan Carlos-Banco de Santander (Excellence Group QUINANOAP, grant number 30VCPIGI14). 

\pagebreak


\begin{thebibliography}{99}
   \bibitem{olv1} P.J. Olver, Applications of Lie Groups to Differential Equations, second ed.,  
		Springer-Verlag, New York, 1993.
   \bibitem{bn4} B. Hern\'{a}ndez-Bermejo, New solution family of the Jacobi equations: Characterization, 	
   		invariants, and global Darboux analysis, J. Math. Phys. 48 022903 (2007) 1--11.
   \bibitem{bma1} B. Hern\'{a}ndez-Bermejo, Generalization of solutions of the Jacobi PDEs associated to 
   		time reparametrizations of Poisson systems, J. Math. Anal. Appl. 344 (2008) 655--666.
	\bibitem{dyh} D. David, D. D. Holm, Multiple Lie-Poisson structures, reductions, and geometric
		phases for the Maxwell-Bloch travelling wave equations, J. Nonlinear Sci. 2 (1992) 241--262.
	\bibitem{pyj} G. Picard, T. W. Johnston, Instability cascades, Lotka-Volterra population equations,
		and Hamiltonian chaos, Phys. Rev. Lett. 48 (1982) 1610--1613.
   \bibitem{sep1} B. Hern\'{a}ndez-Bermejo, V. Fair\'{e}n, Separation of variables in the Jacobi 
		identities, Phys. Lett. 271A (2000) 258--263.
   \bibitem{sep2} B. Hern\'{a}ndez-Bermejo, Generalization of the separation of variables in the 
   		Jacobi identities for finite-dimensional Poisson systems, Phys. Lett. 375A (2011) 1972–-1975. 
   \bibitem{agz1} A. Ay, M. G\"{u}rses, K. Zheltukhin, Hamiltonian equations in $R^3$, J. Math. Phys. 
   		44 (2003) 5688–-5705.
   \bibitem{cpe1} S.A. Charalambides, P.A. Damianou, C.A. Evripidou, On generalized Volterra systems, 
   		J. Geom. Phys. 87 (2015) 86--105.
   \bibitem{cas1} B. Hern\'{a}ndez-Bermejo, V. Fair\'{e}n, Simple evaluation of Casimir invariants in 
   		finite-dimensional Poisson systems, Phys. Lett. 241A (1998) 148--154.
   \bibitem{ayre} F. Ayres Jr., Schaum's Outline of Matrices, McGraw-Hill, New York, 1962.
   \bibitem{frk1} T. Frankel, The Geometry of Physics, second ed., Cambridge University Press, 
		Cambridge UK, 2004.
   \bibitem{kmck} Y. Nutku, Bi-Hamiltonian structure of the Kermack-McKendrick model for epidemics,
		J. Phys. A: Math. Gen. 23 (1990) L1145--L1146.
   \bibitem{dam1} P.A. Damianou, S. Kouzaris, Bogoyavlensky-Toda systems of type $D_N$, J. Phys. A: 
   		Math. Gen. 36 (2003) 1385--1399. 
   \bibitem{dam2} S.A. Charalambides, P.A. Damianou, $so(p, q)$ Toda systems, Physica 248D (2013) 33--43. 
   \bibitem{mm1} A.V. Borisov, I.S. Mamaev, Rolling of a rigid body on plane and sphere. Hierarchy 
   		of dynamics, Regul. Chaotic Dyn. 7 (2002) 177--200.
   \bibitem{mm2} A.V. Borisov, I.S. Mamaev, A.A. Kilin, Rolling of a ball on a surface. New integrals 
   		and hierarchy of dynamics, Regul. Chaotic Dyn. 7 (2002) 201--219.
   \bibitem{ar1} A. Ramos, Poisson Structures for Reduced Non-Holonomic Systems, J. Phys. A: Math. Gen. 
   		37 (2004) 4821--4842.
   \bibitem{fgs1} F. Fass\`{o}, A. Giacobbe, N. Sansonetto, Periodic flows, rank-two Poisson structures, 
   		and nonholonomic mechanics, Regul. Chaotic Dyn. 10 (2005) 267--284.
   \bibitem{ntt2} B. Hern\'{a}ndez-Bermejo, Generalized results on the role of new-time transformations 
   		in finite-dimensional Poisson systems, Phys. Lett. 374A (2010) 836--841.
   \bibitem{agfs} E. Abado\u{g}lu, H. G\={u}mral, Bi-Hamiltonian structure in Frenet-Serret frame, 
   		Physica 238D (2009) 526--530.
   \bibitem{hgdr} J. Hietarinta, B. Grammaticos, B. Dorizzi, A. Ramani, Coupling-Constant 
   		Metamorphosis and Duality between Integrable Hamiltonian Systems, Phys. Rev. Lett. 53 (1984) 
   		1707--1710.
	\bibitem{rgb} A. Ramani, B. Grammaticos, T. Bountis, The Painlev\'{e} property and singularity analysis 
		of integrable and non-integrable systems, Phys. Rep. 180 (1989) 159--245.
   \bibitem{gor1} A. Goriely, Integrability and Nonintegrability of Dynamical Systems, World Scientific, 
   		Singapore, 2001.
   \bibitem{mha} K.R. Meyer, G.R. Hall, Introduction to Hamiltonian Dynamical Systems and the N-Body Problem, 
   		Springer-Verlag, New York, 1992.
   \bibitem{bn0} B. Hern\'{a}ndez-Bermejo, Characterization, global analysis and integrability of a family 
   		of Poisson structures, Phys. Lett. 372A (2008) 1009--1017.
   \bibitem{bn5} B. Hern\'{a}ndez-Bermejo, An integrable family of Poisson systems: Characterization and
		global analysis, Appl. Math. Lett. 22 (2009) 187--191.
   \bibitem{pd14} I. A. Garc\'{\i}a, B. Hern\'{a}ndez-Bermejo, Periodic orbits in analytically perturbed 
   		Poisson systems, Physica 276D (2014) 1--6.
   \bibitem{dist} B. Hern\'{a}ndez-Bermejo, New global solutions of the Jacobi partial differential 
		equations, Physica 241D (2012) 764--774.
\end{thebibliography}
\end{document}